%% file: main.tex
\documentclass[conference]{IEEEtran}
\IEEEoverridecommandlockouts
\usepackage{cite}
\usepackage{amsmath,amssymb,amsfonts}
\usepackage{algorithmic}
\usepackage{graphicx}
\usepackage{textcomp}
\usepackage{xcolor}
\usepackage{float}
\usepackage{wrapfig}
\usepackage{titlesec}
\usepackage{afterpage} 
\def\BibTeX{{\rm B\kern-.05em{\sc i\kern-.025em b}\kern-.08em
    T\kern-.1667em\lower.7ex\hbox{E}\kern-.125emX}}
\begin{document}


\title{MetaDetect: Metamorphic Testing Based Anomaly Detection for Multi-UAV Wireless Networks
}

\author{\IEEEauthorblockN{Boyang Yan}
\IEEEauthorblockA{\textit{Computer Science} \\
\textit{NC State University}\\
Raleigh, US \\
ORCID: 0000-0002-8546-5004}
}

\maketitle

\input{Abstract}


\input{Introduction}
\input{RelatedWorks}

\input{Methods}
\input{Experiments}

\input{Results}
\input{Discussion}
\input{Acknowledgment}
\input{References}

\end{document}

%% file: Abstract.tex
\begin{abstract}
The reliability of wireless Ad Hoc Networks (WANET) communication is much lower than wired networks. WANET will be impacted by node overload, routing protocol, weather, obstacle blockage, and many other factors, all those anomalies cannot be avoided. Accurate prediction of the network entirely stopping in advance is essential after people could do networking re-routing or changing to different bands. In the present study, there are two primary goals. Firstly, design anomaly events detection patterns based on Metamorphic Testing (MT) methodology. Secondly, compare the performance of evaluation metrics, such as Transfer Rate, Occupancy rate, and the Number of packets received. Compared to other studies, the most significant advantage of mathematical interpretability, as well as not requiring dependence on physical environmental information, only relies on the networking physical layer and Mac layer data. The analysis of the results demonstrates that the proposed MT detection method is helpful for automatically identifying incidents/accident events on WANET. The physical layer transfer Rate metric could get the best performance.
\end{abstract}

%% file: Introduction.tex
\section{Introduction}
\label{sec:intro}
The next generation of wireless networks, 6G and beyond, will rely on vertical heterogeneous networks (VHetNets) that integrate space networks (comprising low, medium, and geostationary Earth orbit satellites), air networks (including high-altitude platform stations (HAPSs) and unmanned aerial vehicles (UAVs)), and terrestrial networks (comprising macro and micro base stations) to deliver comprehensive global coverage \cite{reference9}. Unmanned aerial vehicles (UAVs), also known as Autonomous Aerial Vehicles (AAVs), Remotely Operated Aerial Vehicles (RPVs), or unmanned aerial vehicles \cite{reference10}, are a critical component of the urban low-altitude network due to UAVs convenience and flexibility in executing various missions compared to manned aircraft. Wireless networks and UAV communication complement each other. Wireless networks play a vital role in supporting UAV communication, as they offer flexibility, mobility, and scalability, making it possible for UAVs to operate effectively in a wide range of environments and applications. As UAV technology continues to evolve, the importance of wireless networks is only expected to grow.\par

The reliability of wireless network communication (WNC) is much lower than wired networks. WNC will impact by node overload, routing protocol, weather, obstacle blockage, and many other factors, all those anomalies cannot be avoided \cite{reference13}. The purpose of detecting anomalies is defined as finding non-compliance consistent with the concept of normal behavior \cite{reference11}. Other widely used words are outliers, exceptions, surprises, and aberrations \cite{reference12}. The need for action on the relevance of the issue finds outliers. Accurate prediction of the network entirely stopping in advance is essential after people could do networking re-routing or changing to different bands.\par

The next-generation wireless networks are evolving into very complex systems \cite{reference16}. To optimize the performance and detect anomalies of these networks, a more holistic approach is needed, which takes into account the interdependence between different network elements and the diverse requirements of different applications and services. In recent years, scholars have been trying to achieve this through the use of machine learning (ML) and deep neural networks (DNNs) techniques which have been reported with better performance. AI algorithms can learn from data and observations and can be used to identify patterns and make predictions in real-time, based on changing conditions and demands that traditional data analysis tools may miss. Audibert et al. [72] proposed USAD, in which one encoder and two decoders are trained adversarially, to identify network anomalies. This can help network operators to identify potential issues and proactively address them before they become major problems. As these networks continue to evolve and become more complex, network operators will need to embrace these technologies in order to stay competitive and meet the diversity.

However, there is also growing concern about their black-box nature. DNNs are often found to exhibit unexpected behaviors. The interpretability issue affects people’s trust in deep learning systems \cite{reference14}. The definition of interpretability is the ability to explain or to present in understandable terms to a human \cite{reference15}. There are three categorical of evaluation approaches for interpretability: application-grounded, human-grounded, and functionally-grounded \cite{reference15}.
\begin{itemize}
  \item Application-grounded evaluation involves conducting domain expert experiments within a real application.
  \item Human-grounded evaluation is about conducting simpler human-subject experiments that maintain the essence of the target application. Such an evaluation is appealing when experiments with the target community is challenging. These evaluations can be completed with lay humans, allowing for both a bigger subject pool and less expenses, since we do not have to compensate highly trained domain experts.
  \item Functionally-grounded evaluation requires no human experiments; instead, it uses some formal definition of interpretability as a proxy for explanation quality.
\end{itemize}
Those three evaluation approaches are highly costs.

\par

Traditional software has human pre-defined rules and structure. Testers design test suits according to the software specification, and then they test whether the software’s behavior is correct \cite{reference17}. There is a common test oracle problem in large complex traditional software systems. It is difficult to determine the expected outcomes of selected test cases or to determine whether the actual outputs agree with the expected outcomes. For example, the elevation of \(sin\) function. Such as \(sin (30^{\circ}) = 0.5\), we could decide whether the output is correct or not very easily. But, how about \(sin(2.7)\), \(sin(3.4)\). it is difficult to answer whether the output is right or wrong. As we all know, different from traditional software, DNNs-driven software has the following characteristics \cite{reference18}:
\begin{enumerate}
  \item No definite logical structure and specification: DNNs-driven software the structure and logic emerge from the training data. The software consists of a neural network that learns to map inputs to outputs based on patterns in the data.
  \item Huge input and output space: Deep learning models are capable of processing vast amounts of data, and can handle complex inputs such as images, audio, and natural language. The output space can also be large, especially in tasks such as image or speech recognition, where there are many possible outputs.
  \item Uncertainty in output: Because deep learning models learn from data, their output can be uncertain or probabilistic. For example, a deep learning model that is trained to recognize images of cats may assign a certain probability to each image that it is a cat. This uncertainty is often represented as a probability distribution over possible outputs.
\end{enumerate}

As a result, the test oracle problem is a more common issue in the context of DNNs-driven software as it is difficult to determine the correct output of a DNN for a given input. The current research community has extended and innovated on the basis of traditional software testing methods, proposing a series of new testing techniques for DL-driven software, including Metamorphic Testing (MT) \cite{reference36, reference37}, Combinatorial Testing \cite{reference38}, and Fuzzing Testing \cite{reference39}. Relatively, MT is one of the newer and more effective testing techniques that have been proposed for testing DNNs-driven software. MT works by identifying and using the metamorphic relations that hold between inputs and outputs of a DNN. These metamorphic relations are properties that should hold true for any input and output pair that satisfies the relation. By applying the metamorphic relations to a set of inputs and their corresponding outputs, MT can generate new test cases that can reveal potential faults or errors to provide a more comprehensive testing approach for DNNs-driven software.

Metamorphic testing is particularly useful for testing complex systems, such as machine learning algorithms, where it can be difficult to generate a complete set of test cases that cover all possible scenarios. By focusing on the relationships between input and output, rather than on individual inputs and outputs, metamorphic testing can help identify faults and improve the overall quality of the system.

In the present study, there are two primary goals. Firstly, design anomaly events detection patterns based on Metamorphic Testing (MT) methodology. Secondly, compare the performance of evaluation metrics, such as Transfer Rate, Occupancy rate, and the Number of packets received. Compared to other studies, the most significant advantage of mathematical interpretability, as well as not requiring dependence on physical environmental information, only relies on the networking physical layer and Mac layer data. The analysis of the results demonstrates that the proposed MT detection method is helpful for automatically identifying incidents/accident events on WNC. The physical layer transfer Rate metric could get the best performance.

To summarize, this paper makes three major contributions:
\begin{itemize}
  \item We propose applying MT to evaluate wireless anaonly events without using ground truth labels (100\% packet loss).
  \item We design eight MRs for anaonly events detection, each of which focuses on one specific property of network node. These MRs could be understanding the characteristics from a multi-view.
  \item 
\end{itemize}

The rest of this paper is organized as follows: Related Works, Proposed Methods, System Architecture, Evaluation, Discussion, and Conclusion.

%% file: RelatedWorks.tex
\section{Related Works}

\subsection{Metamorphic testing}
Metamorphic testing (MT) is one of the property-based software testing techniques that was originally developed as a software verification method in the Software Engineering Research Area. It belongs to Black-Box Testing. The metamorphic word means changing or transforming from one status to another status, and similar meaning to meta-mutation. The Core Concept of MT is to set up mathematics relations. In conventional metamorphic testing, it called metamorphic relations (MRs) are identified as necessary properties of the software under test (SUT)’s intended functionality against prescribed MRs. An MR is an expected relationship among the inputs and outputs of multiple executions of the SUT. It can be an effective approach for addressing the mechanism for determining whether a test has passed or failed (reveal faults) and test case generation problems [1]. It is especially useful in cases where it is difficult or impossible to generate a complete set of test cases that can cover all possible scenarios.

For example, if we are testing a weather app that provides temperature information for a given location, one metamorphic relation could be that the temperature should be lower at night than during the day. To test this relation, the tester would generate two sets of test cases: one set with daytime input data and one set with nighttime input data. The expected output for the daytime test cases should be higher than the expected output for the nighttime test cases.

In the present study, we argue that MRs can also be used to detect network traffic anomalies. The structure of the programming code is seen as our Wireless Network Topology. In addition, the software test suits are our Transmission packages. Every Transmission Packet's current status, such as position and time, depends on the previous state. Based on this new perspective, we designed three MRs based on different MANET nodes' Statistical information. If the relation is broken, we can say this is a traffic anomaly event that may happen.

We are using, which NS3 discrete-event network simulator to generate the primary dataset of this study. The ground truth data used Nakagami-Rice Distribution as our loss function. However, the ground truth data was used to evaluate the MRs; and accuracy rate, not for designing MRs.

\subsection{Metamorphic Relations for Data Validation}

\includegraphics[width=0.45\textwidth]{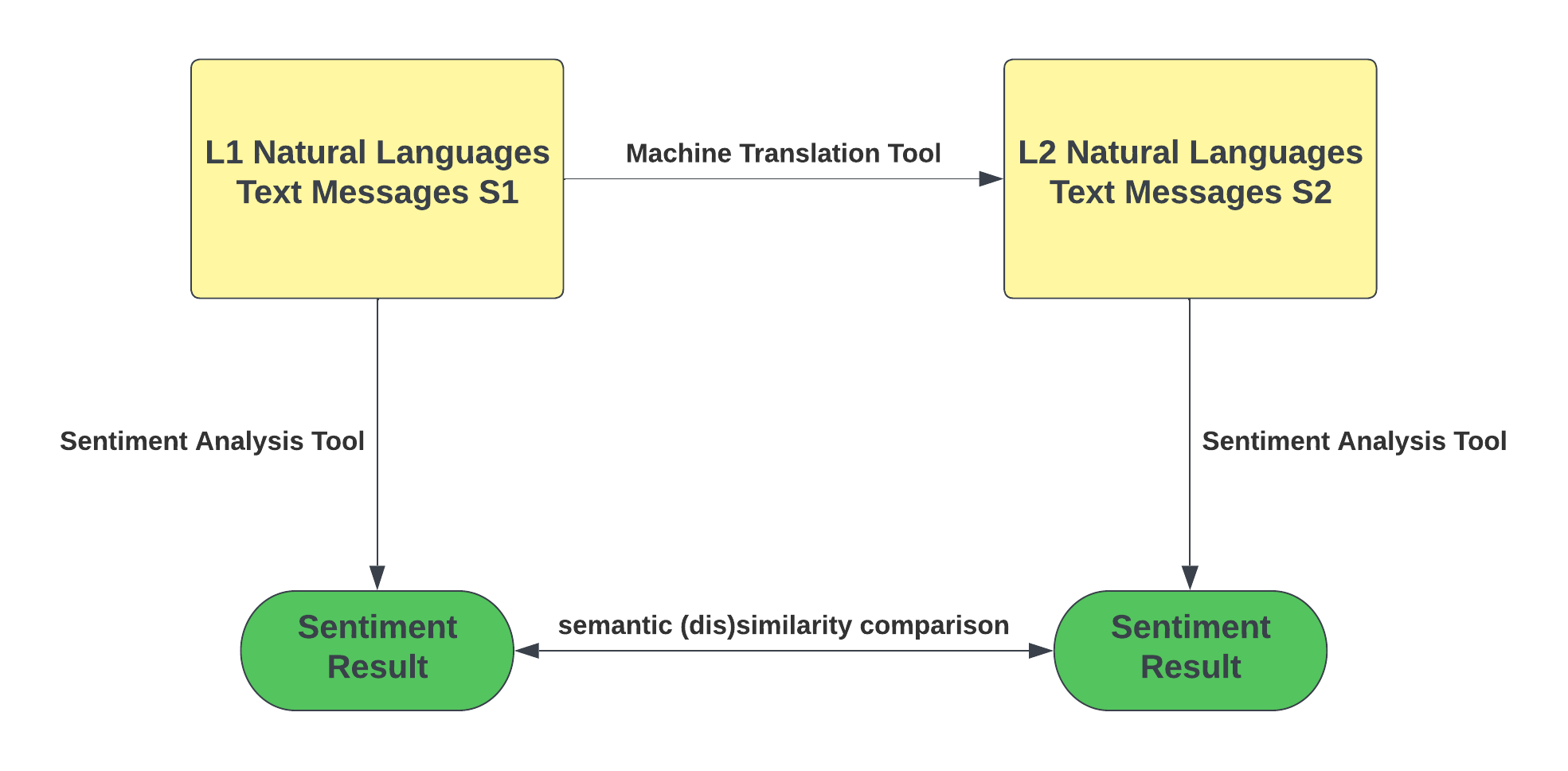}

Machine Translation of translated Text Validation [1]. Let L1 and L2 be two different natural languages. Let s1 and s2 be text messages written in L1 and L2 languages. Let a sentiment analysis tool that can work for both languages, L1 and L2. If s2 is an accurate translation of s1, then the sentiment analysis results produced by the sentiment analysis tool for s1 and s2 should be similar.

In this example, the S1 we call the source test case, and the S2 is the follow-up test case. We can verify whether the actual outputs (sentiment result) produced by the program under test from the source test case and the follow-up test case are consistent with the MR in question. Any inconsistency indicates a failure of the program caused by a fault in the implementation. The intention of designing this MR is not to test the sentiment analysis tool but rather to evaluate the semantic (dis)similarity between s1 and s2: If the sentiment analysis tool is a high-quality sentiment analysis tool, then a violation of MR could indicate that s2 is a poor translation of s1. In this way, MR is used for data validation and quality assessment. 

\subsection{Metamorphic Relations for better understand the characteristics}
MT-based evaluation for RE can not only alleviate the oracle problem but also help to better understand the characteristics of RE models from various viewpoints \cite{reference47}.

%% file: Methods.tex
\section{MetaDetect: Method}
In this section, we introduce our approach of MT-based evaluation for Multi-UAV enabled wireless networks anomaly detection and better understand the Multi-UAV characteristics from various viewpoints. 

We first explain the process of performing MT on Multi-UAV enabled wireless networks and then present the details of the designed MRs.

To evaluate the MetaDetect method proposed in this paper, we first constructed a dataset, which is mainly divided into two categories: the data from different network stack layers, UAV sensor data, mainly used to calculate the distance between drones. Then, we design an efficient data collection system (Data Generation and Data Collection). Finally, we conducted our anomaly detection algorithm on Multi-UAV.
\input{Scenario}
\input{DataCollect}
\input{MR}
\input{AnomalyDetection}

%% file: Scenario.tex
\subsection{Accidents Identification Area / Scenario Overview}
\includegraphics[width=0.45\textwidth]{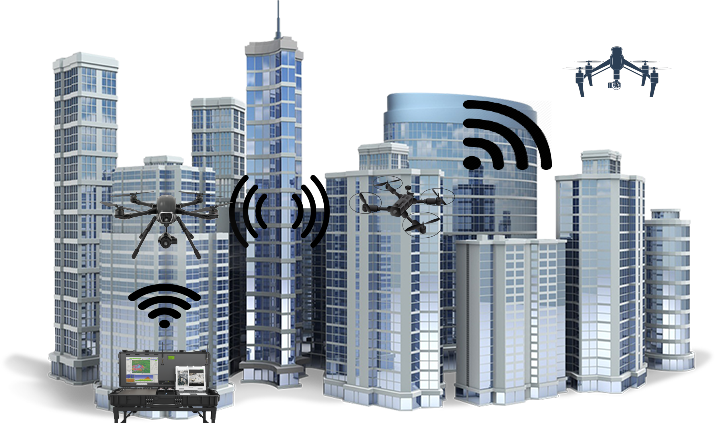}
The dynamics of real-world wireless networks are non-linear and their underlying topology is time-varying \cite{reference19}. However, complex networks with linear dynamics have been intensively researched recently, which can be motivated in several ways \cite{reference20, reference21}. Non-linear dynamics on the networks can be approximated \cite{reference22} or bounded \cite{reference23} by the linear dynamics, in most cases. 

Each Identification Area is a group of four Ad-hoc nodes—for example, one ground station with three UAVs. The simulation is done in NS3. I installed the On-Off server on the last UAV, and you could assume the On-Off server is one of the cameras in the UAV. We want to transfer the video back to the ground station, so I also install a sink application on the Ground station. The routing path is fixed by linear. I used the 3D Gauss Markov Mobility Model for all UAVs; this Mobility Model is the closest to real UAV flying behavior, as mentioned by Li [8]. The ground truth of Anomalies is generated in NS3. Nakamura mentioned that the measured path loss broadly matches the Nakagami-Rice distribution[4], so I used the Nakagami Rice Distribution as our loss function.

\subsection{Workflow performing MT on Multi-UAV}
\includegraphics[width=0.45\textwidth]{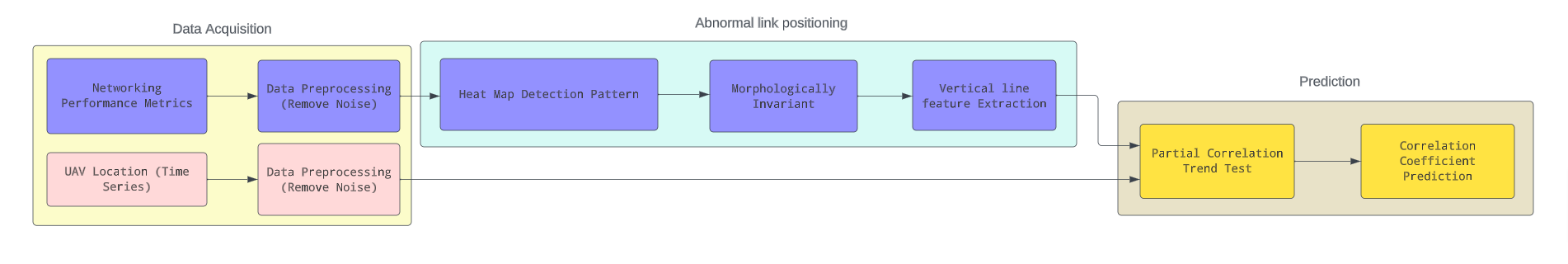}

%% file: DataCollect.tex
\subsection{Data Acquisition}
Based on Yuan et al.'s research on the most significant metrics for assessing wireless network performance, we have extracted eight metrics below \cite{reference43}.

\begin{enumerate}
  \item \textbf{Physical Layer (Phy): Received Signal Strength Indicator (RSSI)} measures the power of the frequency band of interest at the receiver. When there is active transmission during the sampling time, RSSI measures the power of the received signal. In the absence of active transmission, RSSI measures the power of the interference and background noise.
  \item \textbf{Physical Layer (Phy): Link Quality Indicator (LQI)} is used to assess the strength and quality of a received packet. The value of LQI is reported as an integer that should range from 0 to 255. A higher LQI value generally indicates a stronger and more reliable link, while a lower value suggests a weaker or less reliable link.
  \item \textbf{Physical Layer (Phy): Signal to interference and Noise Ratio (SINR)} is calculated using the formula $SNR = \frac{P_{S}}{P_{I+N}}$, where $P_{S}$ represents the power of the signal and $P_{I+N}$ represents the combined power of the interference and background noise.
  \item \textbf{Physical Layer (Phy): Packet Corruption Rate (PCR)} is a metric that calculates the ratio of received corrupted packets to the total number of received packets in a given link.
  \item \textbf{Data Link Layer (MAC): Single-hop delay (SH-Delay)} measures the time it takes for a packet to be fully processed at a receiver after it has been scheduled for transmission at a sender. This metric provides an indication of link congestion and is determined by adding the output queuing and processing delay at the sender ($d_{output}$), propagation delay ($d_{prop}$), transmission delay ($d_{tx}$), and input queuing and processing delay at the receiver ($d_{input}$). The formula for calculating single-hop delay is as follows: $$d = d_{output} + d_{prop} + d_{tx} + d_{input}$$
  \item \textbf{Data Link Layer (MAC): Single-hop jitter (SH-Jitter)} is a metric that calculates the variation in single-hop delay.
  \item \textbf{Data Link Layer (MAC): Single-hop throughput (SH-Throughput)} measures the successful delivery of packet payload over a link within a specified time period. In the absence of interference from other concurrently active links and noise, the single-hop throughput should approach the maximum attainable data rate of the link. Additionally, higher link quality typically results in higher single-hop throughput, as fewer packet retransmissions are needed to deliver the packet over the link.
  \item \textbf{Data Link Layer (MAC): Single-hop packet reception rate (SH-PRR)} is the most commonly used metric for evaluating link quality. Packet reception rate (PRR) represents the percentage of packets that are successfully delivered over a link. PRR is sometimes referred to as packet delivery ratio (PDR), and its complement is known as packet loss rate (PLR).
  \item \textbf{Data Link Layer (MAC): $\beta$ factor} characterizes the burstiness of a wireless link. It is a scalar value, with 0 being an independent link and 1 being a perfectly bursty link. The $\beta$ factor is derived from the conditional probability delivery function of the measured link. $\beta = \frac{KW(I) - KW(E)}{KW(I)}$, where $I$ is an independent link with the same PRR as the measured link and $E$ is the measured link. $KW (\cdot)$ is the Kantorovich–Wasserstein distance of a link to the perfectly bursty link. The KW distance is defined as the average of the absolute differences of the corresponding elements of two vectors. It is shown that with increasing inter-packet interval, $\beta$ comes close to 0.
\end{enumerate}

Even Mini-UAV (below 20kg) \cite{reference41} could have multiple composite positioning methods (such as GPS+RTK, optical flow, and inertial navigation), and it could get robust location results.

Nancy et al. conducted an in-depth evaluation of different distance metrics and proposed the most appropriate combination of metrics for different situations. Generally, the Euclidean metric is a good choice \cite{reference42}. Wireless propagation is always linear, so the Euclidean metric will be the best option.

$$d = \sqrt{(x_2 - x_1)^2 + (y_2 - y_1)^2 + (z_2 - z_1)^2}$$

In this formula, (x1, y1, z1) and (x2, y2, z2) are the coordinates of two points in 3D space, and d is the distance between them. The formula calculates the distance between the two points by taking the square root of the sum of the squares of the differences between their x, y, and z coordinates.

%% file: MR.tex
\subsection{Metamorphic Relations}

We totally designed Nine MRs, based on nine evaluation metrics.

Partial correlation trend test
This test performs a partial correlation trend test with either the Pearson's or the Spearman's correlation coecients $$(r(tx.z))$$. The magnitude of the linear / monotonic trend
with time is computed while the impact of the co-variate is partialled out.

\subsubsection{MR1: RSSI}
The distance between the two UAVs have decreased, the RSSI also has decreased, which indicating an anomaly event.

$$RSSI_{TS} = (RSSI_{t}: t \in T)$$
$$Distance_{TS} = (Distance_{t}: t \in T)$$
where $T$ is the time index set.

$$corr(RSSI_{TS}, Distance_{TS}) > 0$$

\subsubsection{MR2: LQI}
The distance between the two UAVs have decreased, the LQI also has decreased, which indicating an anomaly event.

\subsubsection{MR3: SINR}
The distance between the two UAVs have decreased, the SINR has increased, which indicating an anomaly event.

\subsubsection{MR4: PCR}
The distance between the two UAVs have decreased, the PCR has increased, which indicating an anomaly event.

\subsubsection{MR5: SH-Delay}
The distance between the two UAVs have decreased, the SH-Delay has increased, which indicating an anomaly event.

\subsubsection{MR6: SH-Jitter}
Change in distance between the two UAVs is negatively correlated with change in SH-Jitter, which indicating an anomaly event.

\subsubsection{MR7: SH-Throughput}
The distance between the two UAVs have decreased, the SH-Throughput has decreased, which indicating an anomaly event.

\subsubsection{MR8: SH-PRR}
The distance between the two UAVs have decreased, the SH-PRR has decreased, which indicating an anomaly event.

\subsubsection{MR9: $\beta$ factor }
The distance between the two UAVs have decreased, the $\beta$ factor change from 1 to 0, which indicating an anomaly event.

Through the above 9 MRs, we can distinguish whether anomalies can are detected in Accidents Identification Area; there are two new problems be raised.
\begin{enumerate}
  \item Abnormal link positioning
  \item Future anomaly event prediction based on evaluation metrics 
\end{enumerate}

%% file: AnomalyDetection.tex
\subsection{Anomaly Detection Model}

\subsubsection{Data Preprocessing}
Data Preprocessing, which is also commonly referred to as data wrangling, data cleaning, etc., is a process and the collection of operations needed to prepare all forms of untidy data (incomplete, noisy and inconsistent data) for statistical analysis. In practice, a data analyst spends most of time (usually 50\%-80\% of an analyst time) on making ready the data before doing any statistical operation \cite{reference44}

\paragraph{Data Interpretation}
We cannot perform any type of data preprocessing without understanding what we have in hand. In this step, the most importance is understand the types of metrics and levels of measurement.

\vspace{+0.2cm}
\begin{tabular}{ |p{0.2\textwidth}||p{0.2\textwidth}| }
 \hline
 \multicolumn{2}{|c|}{Types of Metrics and Levels of Measurement} \\
 \hline
 Metrics Name & Types-Levels\\
 \hline
 Phy-RSSI & Numerical-Ratio\\
 Phy-LQI & Numerical-Interval\\
 Phy-SNR & Numerical-Ratio\\
 Phy-PCR & Numerical-Ratio\\
 Mac-SH-Delay & Numerical-Ratio\\
 Mac-SH-Jitter & Numerical-Ratio\\
 MAC-SH-Throughput & Numerical-Ratio\\
 MAC-SH-PRR & Numerical-Ratio\\
 MAC-$\beta$ factor & Categorical-Nominal\\
 UAV-Distance & Numerical-Ratio\\
 \hline
\end{tabular}

\paragraph{Data Imputation}
Although Data Collection become more reliable, abnormal data are found in the database due to communication failure, device malfunction, or shut down by users. Missing values and unreasonably excessive loads are considered abnormal data.

\vspace{+0.2cm}
\begin{tabular}{ |p{0.2\textwidth}||p{0.2\textwidth}| }
 \hline
 \multicolumn{2}{|c|}{Data Imputation} \\
 \hline
 Metrics Name & Imputation Value\\
 \hline
 Phy-RSSI & 0\\
 Phy-LQI & 0\\
 Phy-SNR & 0\\
 Phy-PCR & 100\%\\
 Mac-SH-Delay & Positive Infinity\\
 Mac-SH-Jitter & Positive Infinity\\
 MAC-SH-Throughput & 0\\
 MAC-SH-PRR & 0\\
 MAC-$\beta$ factor & 0\\
 UAV-Distance & Positive Infinity\\
 \hline
\end{tabular}

\paragraph{Multi-dimension Data Transformation}
After data imputation, data transformation must be performed before statistical analysis/modeling to provide patterns that are easier to understand. Data transformation changes the format, structure, or values of the data and converts them into clean, usable data. It increases the efficiency of analytic processes and enables better data-driven decisions. Several data transformation techniques can help structure and clean up the data, such as Data Smoothing, Attribute Construction, Data Aggregation, Data Normalization, Data Discretization, and Data Generalization.

UAV Distance:
UAVs impossible flash move from one location to another, the change of distance is continuous. In UAV smooth motion planning, the cubic spline method is the most used to fit the path point with the corresponding time interval. Using cubic spline and time reallocation mechanism, we can obtain a smooth trajectory that satisfies the dynamic constraints\cite{reference45}.

wireless measurement metrics:
The utilization and statistical methods used for these metrics can vary significantly, leading to failure in describing their capabilities \cite{reference46}. In some cases, different studies may even reach contradicting conclusions regarding the effectiveness of certain metrics.

In this paper, Phy-RSSI, Phy-LQI, Phy-SNR, Phy-PCR

\subsubsection{Abnormal link positioning (Heat Map Detection method/pattern)}
\includegraphics[width=0.45\textwidth]{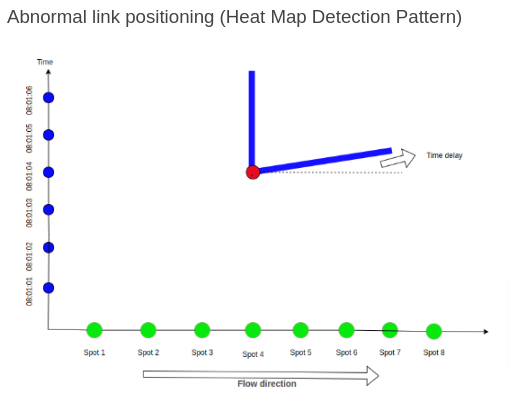}
The evaluation metrics value will be the x-axis. The axis direction is the same as the network traffic direction, which means the direction from left to right. The y-axis represents the time because all the data collected are time-series data. The heat map will show a semi-triangle with only two edges if an anomaly is detected. The vertex of the semi-triangle, the corresponding y-axis, is when the abnormality happened, and the corresponding x-axis is which node has the anaonly event.

\paragraph{Morphologically Invariant}

Morphologically invariant transformations represent a crucial concept in the field of computer vision, particularly for enhancing robustness against significant illumination changes. These transformations are characterized by their extraordinary ability to remain consistent under any global, monotonically increasing rescaling of the input signal. This characteristic is referred to as morphological invariance. Such invariance ensures that the transformed signal maintains its essential features regardless of variations in intensity, thereby enabling algorithms to perform consistently in diverse lighting conditions. Morphologically invariant transformations are designed to encapsulate the maximum amount of information possible, which is pivotal for accurate and reliable matching of structures in computer vision applications. Among the algorithms that leverage this property, the rank transform and census transform are the most prominent. Both algorithms utilize the principle of morphological invariance to efficiently handle variations in illumination, thus playing a significant role in the development of robust computer vision systems.

\paragraph{Vertical line feature Extraction}
Vertical line feature extraction is an essential technique in the realm of image processing and computer vision, facilitating the identification and analysis of vertical structures within a digital image. This process involves the implementation of algorithms that can discern vertical lines from other elements in the image. Typically, edge detection methods, such as the Sobel or Canny algorithms, are employed to highlight areas of high-intensity gradient, which often correspond to the edges of objects. Subsequently, these detected edges are analyzed using techniques like the Hough Transform, which is particularly effective in isolating features with specific orientations, such as vertical lines.

\subsubsection{Future anomaly event prediction}

ARMA + LSTM and Distance Correlation:
The partial correlation trend test is a statistical method used to identify the presence of trends in time series data while controlling for the influence of one or more confounding variables. The ability of the partial correlation trend test to account for confounding factors makes it an indispensable tool in time series analysis, providing insights that are more nuanced and reliable than those obtained from simple correlation or regression analyses.

Pearson’s or the Spearman’s correlation coefficients

%% file: Experiments.tex
\subsection{Evaluation Results}
To evaluate the MetaDetect method proposed in this paper, we first constructed a dataset, which is mainly divided into two categories: the data from different network stack layers, and UAV sersor data, mainly used to calculate the distance between drones. Then, we design an efficient data collection system (Data Generation and Data Collection). Finally, we conducted our anomaly detection algorithm on Multi-UAV.

\input{SystemArchitecture}

%% file: SystemArchitecture.tex
\subsubsection{System Architecture}
\begin{figure*}[h] 
  \centering
  \includegraphics[width=1.0\textwidth]{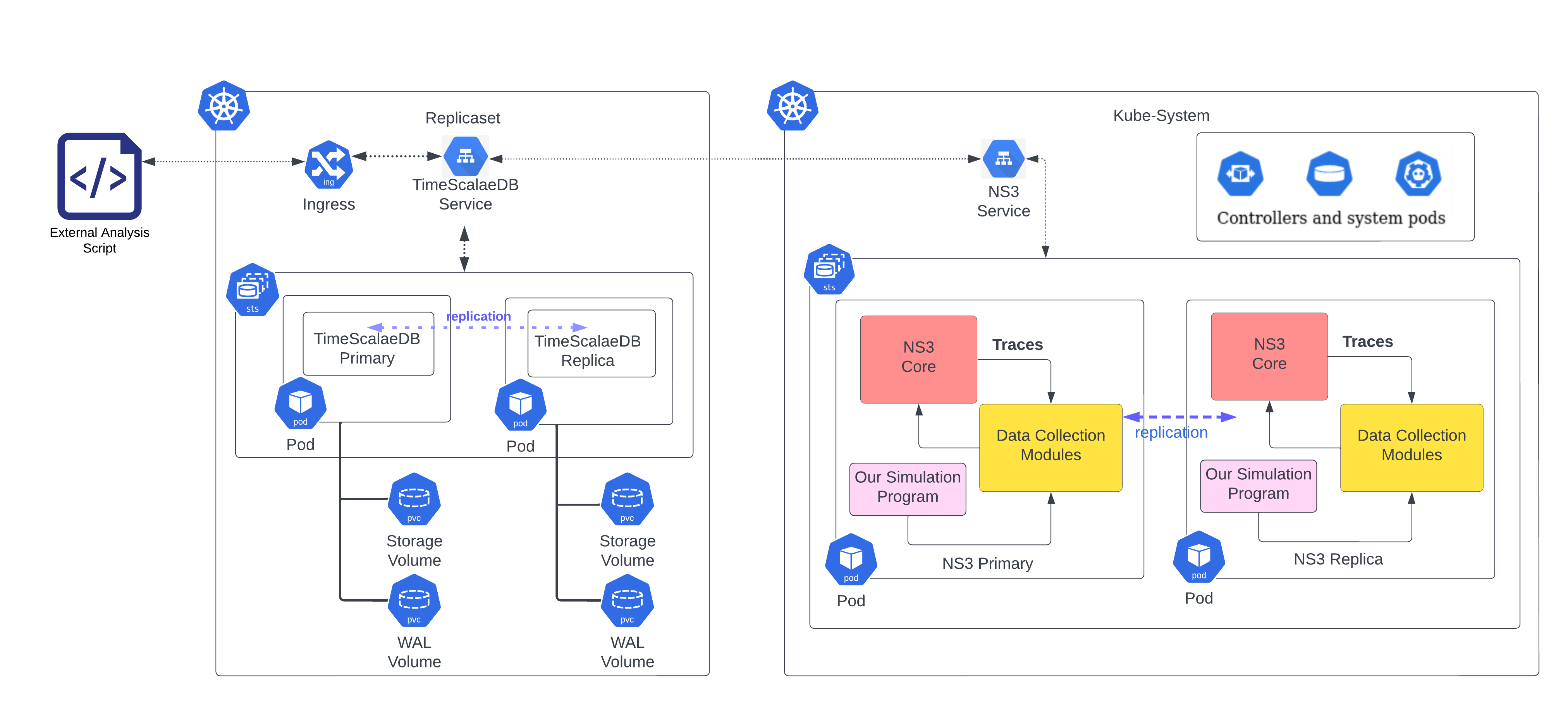}
  \caption{System Architecture}
  \label{fig:SystemArchitecture}
\end{figure*}

Kubernetes (also called K8S), the container orchestrator that has become the most popular cloud manager in the past few years, offers automatic scaling features called Horizontal and Vertical Pod Autoscaling (HPA \cite{reference24}, VPA \cite{reference25}). Those basic components of the System Architecture are shown in figure \ref{fig:SystemArchitecture}, page \pageref{fig:SystemArchitecture} mainly based on this auto-scaling feature to solve the two problems:

\begin{enumerate}
  \item Rapid generation of a large amount of network data
  \item High throughput time-series data storage and SQL query
\end{enumerate}

The System includes two major parts, a time-series database and an NS3 Generator.

Time-Series Database:
TimescaleDB is an open-source Time Series database. It is an extension of PostgreSQL that optimizes for time series data storage and processing. As an extension of PostgreSQL, it is a fully relational database. It provides high levels of analytics abilities with SQL. TimescaleDB also benefits from the PostgreSQL ecosystem of a wide array of extensions, management tools, and visualization dashboards. TimescaleDB transforms PostgreSQL tables into hypertables and optimizes query planning and execution. Hypertables can also be easily deployed in a distributed environment and store data across multiple nodes \cite{reference28}, which could be managed by Kubernetes. TimescaleDB also includes hyperfunctions, which is a set of functions oriented to time series data analysis operations. They extend SQL to achieve fast execution time on time series queries.

Compared with PostgreSQL, TimescaleDB has shown to be 20x faster at inserting large-scale data \cite{reference29}. TimescaleDB can achieve 1.2x to 5x better performance in time-based queries and 450x better performance in time-ordering queries \cite{reference30}. Although InfluxDB is the current most popular Time Series database, TimescaleDB is tested to perform high cardinality inserts 3.5x faster and outperforms InfluxDB by 3.4x to 71x on complex queries \cite{reference31}.

NS3 Generator:
Simulations are a popular methodology for the test and validation of design, models, and systems in a virtual environment under specific situations without having to put the actual corresponding complex system at risk. These simulations carried out through simulation software, tend to be much more realistic and accurate in their outcomes compared to basic mathematical or analytical models \cite{reference32}. Simulation Software is the best supplement to the real environment.

Today, there are several competitive as well as feature-rich network simulators available in the market for researchers, each with its strengths and merits \cite{reference33}. Many of these are capable of not only modeling traditional wired networks but also cutting-edge technologies such as Wi-Fi and 5G networks \cite{reference33}.

NS-3 is a popular discrete-event network simulation software, targeted primarily for research and educational use. NS-3 provides a complex and rich library for the simulation of wireless and wired technology. There are two main benefits:
\begin{itemize}
  \item It supports the integrations with other open-source software, thus reducing the requirement for rewriting simulation models.
  \item It can be used to simulate and analyze large-scale network simulations very efficiently \cite{reference34, reference35}. The protocols are designed to closely model realistic scenario computations and devices.
\end{itemize}

%% file: Results.tex
\section{Evaluation Results}
From the two heat maps below, we could find different Metamorphic Relationships have different performances for finding the network traffic anomaly. MR (phy throughput) gets the best performance and no delay. It shows a strong contrast between normal traffic and traffic anomaly. However, MR (Queue Drop) could detect the traffic anomaly but have a delay, so it could not be used for the forecast.
\includegraphics[width=0.5\textwidth]{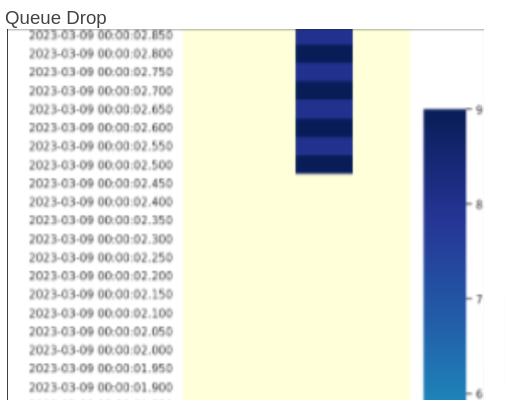}
\includegraphics[width=0.5\textwidth]{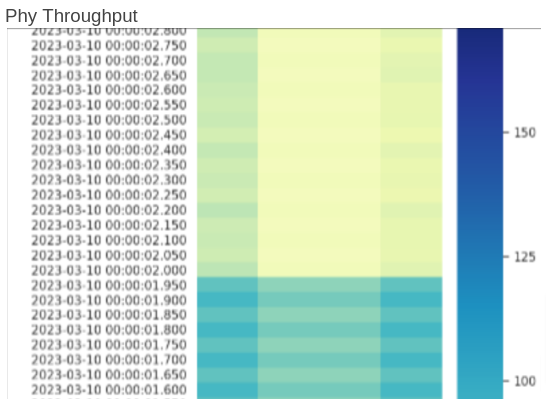}

%% file: Discussion.tex
\section{Conclusion and Discussion}
There are also many other methods based on machine learning(ML) and deep learning(DL) for Wireless Anaonly Detection. Most of those methods are black boxes, which could not explain why. The main advantage of the proposed method (MT) is explainable, which relies entirely on mathematical relations to determine the direction of training and achieve rapid detection. However, the most significant limitation in the current progress of this project is that the Heat Map detection pattern needs manually to distinguish areas of high contrast by humans. It will focus on using a Convolutional Neural Network (CNN, or ConvNet) and recurrent neural network (RNN) to automatically find the areas of significant contrast in the Heat Map; it will be my future work. When the ML/DL algorithm is combined with MT could achieve more reliability and meanwhile resolve some uncertainty factors.

%% file: Acknowledgment.tex
\section{Acknowledgment}
I have not done all parts of this project, but I put all of my current writing and results in Arxiv for the NC State University's CSC801 course report. I will continue doing this project.

%% file: References.tex
\section{References List}

%% file: main.bbl
\begin{thebibliography}{00}

\bibitem{reference1} Yan, B., Yecies, B., Zhou, Z. Q. (2019, May). Metamorphic relations for data validation: A case study of translated text messages. In 2019 IEEE/ACM 4th International Workshop on Metamorphic Testing (MET) (pp. 70-75). IEEE.

\bibitem{reference2} Chandola, V., Banerjee, A., Kumar, V. (2009). Anomaly detection: A survey. ACM computing surveys (CSUR), 41(3), 1-58.

\bibitem{reference3} Emmott, A. F., Das, S., Dietterich, T., Fern, A., Wong, W. K. (2013, August). Systematic construction of anomaly detection benchmarks from real data. In Proceedings of the ACM SIGKDD workshop on outlier detection and description (pp. 16-21).

\bibitem{reference4} Nakamura, M., Sasaki, M., Inomata, M. and Onizawa, T., 2016, October. The effect of human body blockage to path loss characteristics in crowded areas. In 2016 International Symposium on Antennas and Propagation (ISAP) (pp. 218-219). IEEE.

\bibitem{reference5} Burns, W.R., 1964. Some statistical parameters related to the Nakagami-Rice probability distribution. Radio Science, 68, pp.429-34.

\bibitem{reference6} Perkins, D.D., Hughes, H.D. and Owen, C.B., 2002, April. Factors affecting the performance of ad hoc networks. In 2002 IEEE International Conference on Communications. Conference Proceedings. ICC 2002 (Cat. No. 02CH37333) (Vol. 4, pp. 2048-2052). IEEE.

\bibitem{reference7} M. S. Corson, J. P. Macker, and G. H. Cirincione, “Internet-based mobile ad hoc networking”. IEEE Internet Computing, Vol. 3, No. 4, July/August 1999, pp.63-70.

\bibitem{reference8} Li, Y., Wang, W., Gao, H., Wu, Y., Su, M., Wang, J. and Liu, Y., 2020. Air-to-ground 3D channel modeling for UAV based on Gauss-Markov mobile model. AEU-International Journal of Electronics and Communications, 114, p.152995.

\bibitem{reference9} Wang, W., Abbasi, O., Yanikomeroglu, H., Liang, C., Tang, L., \& Chen, Q. (2022). VHetNets for AI and AI for VHetNets: An Anomaly Detection Case Study for Ubiquitous IoT. arXiv preprint arXiv:2210.08132.

\bibitem{reference10} Titouna, C., Naït-Abdesselam, F., \& Moungla, H. (2020, June). An online anomaly detection approach for unmanned aerial vehicles. In 2020 International Wireless Communications and Mobile Computing (IWCMC) (pp. 469-474). IEEE.

\bibitem{reference11} Chandola, V., Banerjee, A., Kumar, V. (2009). Anomaly detection: A survey. ACM computing surveys (CSUR), 41(3), 1-58.

\bibitem{reference12} Emmott, A. F., Das, S., Dietterich, T., Fern, A., Wong, W. K. (2013, August). Systematic construction of anomaly detection benchmarks from real data. In Proceedings of the ACM SIGKDD workshop on outlier detection and description (pp. 16-21).

\bibitem{reference13} Perkins, D. D., Hughes, H. D.,  Owen, C. B. (2002, April). Factors affecting the performance of ad hoc networks. In 2002 IEEE International Conference on Communications. Conference Proceedings. ICC 2002 (Cat. No. 02CH37333) (Vol. 4, pp. 2048-2052). IEEE.

\bibitem{reference14} Zhang, Y., Tiňo, P., Leonardis, A., \& Tang, K. (2021). A survey on neural network interpretability. IEEE Transactions on Emerging Topics in Computational Intelligence, 5(5), 726-742.

\bibitem{reference15} Doshi-Velez, F., \& Kim, B. (2017). Towards a rigorous science of interpretable machine learning. arXiv preprint arXiv:1702.08608.

\bibitem{reference16} Kibria, M. G., Nguyen, K., Villardi, G. P., Zhao, O., Ishizu, K., \& Kojima, F. (2018). Big data analytics, machine learning, and artificial intelligence in next-generation wireless networks. IEEE access, 6, 32328-32338.

\bibitem{reference17} Lo, D., Khoo, S. C., \& Liu, C. (2007, August). Efficient mining of iterative patterns for software specification discovery. In Proceedings of the 13th ACM SIGKDD international conference on Knowledge discovery and data mining (pp. 460-469).

\bibitem{reference18} Zhang, Z., Wang, P., Guo, H., Wang, Z., Zhou, Y., \& Huang, Z. (2021). DeepBackground: Metamorphic testing for Deep-Learning-driven image recognition systems accompanied by Background-Relevance. Information and Software Technology, 140, 106701.

\bibitem{reference19} Strogatz, S. H. (2001). Exploring complex networks. nature, 410(6825), 268-276.

\bibitem{reference20} Xiang, L., Chen, F., Ren, W., \& Chen, G. (2019). Advances in network controllability. IEEE Circuits and Systems Magazine, 19(2), 8-32.

\bibitem{reference21} Liu, Y. Y., \& Barabási, A. L. (2016). Control principles of complex systems. Reviews of Modern Physics, 88(3), 035006.

\bibitem{reference22} Yan, G., Vértes, P. E., Towlson, E. K., Chew, Y. L., Walker, D. S., Schafer, W. R., \& Barabási, A. L. (2017). Network control principles predict neuron function in the Caenorhabditis elegans connectome. Nature, 550(7677), 519-523.

\bibitem{reference23} Prasse, B., \& Van Mieghem, P. (2019). The viral state dynamics of the discrete-time NIMFA epidemic model. IEEE Transactions on Network Science and Engineering, 7(3), 1667-1674.

\bibitem{reference24} Horizontal Pod Autoscaler, Kubernetes, Mountain View, CA, USA. 2020. [Online]. Available: https://kubernetes.io/docs/tasks/run-application/horizontal-pod-autoscale/

\bibitem{reference25} Vertical Pod Autoscaler—Kubernetes. Accessed : Jan. 29, 2021. [Online]. Available: https://github.com/kubernetes/autoscaler/tree/master/verticalpod-autoscaler

\bibitem{reference26} Inc. T. Time-series data simplified — Timescale. 2022. Available from: https://www.timescale.com/

\bibitem{reference28} Inc. T. Distributed hypertables — Timescale Docs. 2022. Available from: https://docs.timescale.com/api/latest/distributed-hypertables/

\bibitem{reference29} Kiefer R. TimescaleDB vs. PostgreSQL for time-series: 20x higher inserts, 2000x faster deletes, 1.2x-14,000x faster queries. 2017. Available from: https://www.timescale.com/blog/timescaledb-vs-6a696248104e/

\bibitem{reference30} Barez, F., Bilokon, P., \& Xiong, R. (2023). Benchmarking Specialized Databases for High-frequency Data. arXiv preprint arXiv:2301.12561.

\bibitem{reference31} InfluxData. influxdb: open source time series database. 2022. Available from: https://www.influxdata.com/products/influxdb-overview/

\bibitem{reference32} Walia, A. K., Chhabra, A., \& Sharma, D. (2022). Comparative Analysis of Contemporary Network Simulators. In Innovative Data Communication Technologies and Application: Proceedings of ICIDCA 2021 (pp. 369-383). Singapore: Springer Nature Singapore.

\bibitem{reference33} Walia, A. K., Chhabra, A., \& Sharma, D. (2022). Comparative Analysis of Contemporary Network Simulators. In Innovative Data Communication Technologies and Application: Proceedings of ICIDCA 2021 (pp. 369-383). Singapore: Springer Nature Singapore.

\bibitem{reference34} Bilalb, S. M., \& Othmana, M. (2013). A performance comparison of network simulators for wireless networks. arXiv preprint arXiv:1307.4129.

\bibitem{reference35} Pan, J., \& Jain, R. (2008). A survey of network simulation tools: Current status and future developments. Email: jp10@ cse. wustl. edu, 2(4), 45.

\bibitem{reference36} Chen, T. Y., Kuo, F. C., Liu, H., Poon, P. L., Towey, D., Tse, T. H., \& Zhou, Z. Q. (2018). Metamorphic testing: A review of challenges and opportunities. ACM Computing Surveys (CSUR), 51(1), 1-27.

\bibitem{reference37} Zhou, Z. Q., \& Sun, L. (2019). Metamorphic testing of driverless cars. Communications of the ACM, 62(3), 61-67.

\bibitem{reference38} Nie, C., \& Leung, H. (2011). A survey of combinatorial testing. ACM Computing Surveys (CSUR), 43(2), 1-29.

\bibitem{reference39} Guo, J., Jiang, Y., Zhao, Y., Chen, Q., \& Sun, J. (2018, October). Dlfuzz: Differential fuzzing testing of deep learning systems. In Proceedings of the 2018 26th ACM Joint Meeting on European Software Engineering Conference and Symposium on the Foundations of Software Engineering (pp. 739-743).

\bibitem{reference40} Nakamura, M., Sasaki, M., Inomata, M., \& Onizawa, T. (2016, October). The effect of human body blockage to path loss characteristics in crowded areas. In 2016 International Symposium on Antennas and Propagation (ISAP) (pp. 218-219). IEEE.

\bibitem{reference41} Singhal, G., Bansod, B., \& Mathew, L. (2018). Unmanned aerial vehicle classification, applications and challenges: A review.

\bibitem{reference42} Amato, N. M., Bayazit, O. B., Dale, L. K., Jones, C., \& Vallejo, D. (1998, May). Choosing good distance metrics and local planners for probabilistic roadmap methods. In Proceedings. 1998 IEEE International Conference on Robotics and Automation (Cat. No. 98CH36146) (Vol. 1, pp. 630-637). IEEE.

\bibitem{reference43} Yuan, D., Kanhere, S. S., \& Hollick, M. (2017). Instrumenting Wireless Sensor Networks—A survey on the metrics that matter. Pervasive and Mobile Computing, 37, 45-62.

\bibitem{reference44} Dasu, T., \&amp; Johnson, T. (2003). Exploratory data mining and data cleaning. Wiley-Interscience.

\bibitem{reference45} Sun, Y., Zhang, C., Sun, P., \& Liu, C. (2020). Safe and smooth motion planning for mecanum-wheeled robot using improved RRT and cubic spline. Arabian Journal for Science and Engineering, 45, 3075-3090.

\bibitem{reference46} Liu, W., Xia, Y., Xu, J., Hu, S., \& Luo, R. (2019, December). Revisiting link quality metrics for wireless sensor networks. In 2019 IEEE 5th International Conference on Computer and Communications (ICCC) (pp. 597-603). IEEE.

\bibitem{reference47} Sun, Y., Ding, Z., Huang, H., Zou, S., \& Jiang, M. (2023). Metamorphic Testing of Relation Extraction Models. Algorithms, 16(2), 102.
\end{thebibliography}
